\documentclass[letterpaper,prl,aps,twocolumn,showpacs]{revtex4}

\usepackage{graphicx}
\usepackage{array}
\usepackage{amsfonts}
\usepackage{amsmath}

\begin{document}

\title{The ultra--sensitive electrical detection of \\ spin Rabi oscillation at paramagnetic defects}
\author{Christoph Boehme\footnote{electronic mail: boehme@hmi.de}, Klaus Lips}
\affiliation{Hahn--Meitner--Institut Berlin, Kekul\'estr. 5, D-12489 Berlin, Germany}
\date{\today}


\begin{abstract}
A short review of the pulsed electrically detected magnetic resonance (pEDMR) experiment is presented. PEDMR allows the highly sensitive
observation of coherent electron spin motion of charge carriers and defects in semiconductors by means of transient current measurements. The
theoretical foundations, the experimental implementation, its sensitivity and its potential with regard to the investigation of electronic
transitions in semiconductors are discussed. For the example of the $\mathrm{P_b}$ center at the crystalline silicon (111) to silicon dioxide
interface it is shown experimentally how one can detect spin Rabi-oscillation, its dephasing, coherence decays and spin--coupling effects .
\end{abstract}


\maketitle

\section{Introduction}
Electron spin resonance (ESR) has proven in the past to be a useful characterization method for the microscopic investigation of paramagnetic
semiconductor defects. The limitations of ESR spectroscopy on semiconductors is set by its sensitivity. Wavelengths in the microwave range are too
long to be detected as single photons. Thus, as low dimensional semiconductors and mesoscopic structures such as quantum--wells, -dots or -wires
or semiconductor thin films have increasingly become subjects of research, ESR spectroscopy that typically reaches sensitivity limits of the order
of $10^{11}$ spins for semiconductor samples at the widely used X-Band ($\approx 10$GHz), is hardly applicable anymore.

In order to achieve higher sensitivities, magnetic resonance methods have been combined in the past with other measurement techniques such as
force microscopy~\cite{Rugar:2004}, photoluminescence~\cite{Jelezko:2002,Jelezko:2004} or conductivity measurements~\cite{Kouven:2004,Jiang:2004}
which have all reached single spin detection sensitivity in recent years. Among these methods, the electrically detected magnetic resonance (EDMR)
technique may be most beneficial for the spin spectroscopy of semiconductors since naturally, it is very sensitive to centers which influence
conductivity while it is blind to all other spins. For most of the EDMR studies found in literature, including those reporting  on a single spin
detection, the experiments were conducted as pure continuous wave (cw) measurements; Pulsed (p) EDMR experiments have only been demonstrated
recently~\cite{Boediss,Boe7}. Since pEDMR combines the advantages of pulsed ESR with those of cw EDMR, these first results suggest that new
insights can potentially be found for the many materials on which cw EDMR has been performed in the past. Beyond material spectroscopy, pEDMR is
also expected to play a role for semiconductor based quantum information concepts: So far, only electrical single spin detection but not
electrical single spin readout experiments have been demonstrated~\cite{Kouven:2004,Jiang:2004}. A readout of a single spin requires a coherent
spin measurement that allows to distinguish between different eigenstates. A requirement which can only be met by coherent spin--detection schemes
as used for pEDMR experiments.

In the following, a brief review of the theoretical and experimental foundations of pEDMR experiments is given. It is shown how one can access
coherent spin--Rabi oscillation by means of electric currents and how this observation can reveal insights into the nature of the observed defect
centers. The sensitivity limitations of pEDMR are addressed, too. As model system, $\mathrm{P_b}$ centers located at the crystalline silicon
(c-Si) (111)/silicon dioxide ($\mathrm{SiO_2}$) interface are used.

\section{Theoretical background} \label{theopart}
PEDMR takes advantage of the spin dependency of charge carrier transitions in semiconductors which occurs when spin conservation is imposed on
electronic transitions. Spin--dependent transition rates between paramagnetic centers can be described in terms of a spin pair ensemble $\hat\rho$
consisting of pairs of two spins with $s=1/2$ corresponding to the two states between which transitions occur~\cite{KSM,HabDie,Boe6}. The
transition rates will be proportional to the singlet content $\mathbf{Tr}\left[|S\rangle\langle S|\hat\rho\right]$ and thus, by measuring currents
as a function of time, the evolution of $\hat\rho$ can be accessed.
\begin{figure}[t]
\begin{center}
\includegraphics[width=84mm]{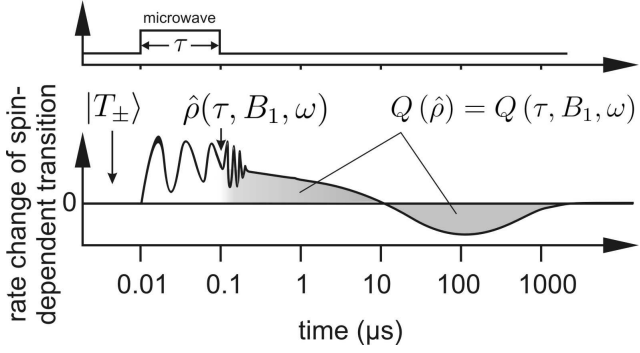}
\caption{The measurement principle of pEDMR on a logarithmic time scale. The integration of the transition rate
$Q$ after a coherent spin excitation is proportional to the singlet content of the ensemble state $\rho(\tau)$ at the end of the pulse. For
details see text.} \label{fig:overview}
\end{center}
\end{figure}
The challenge for pEDMR measurements is to detect very small current changes with high time resolution on top of comparatively large constant
current offsets. It is usually impossible to attain a time resolution with electrical measurements that is within the coherence time of the spin
systems and that is at the same time sensitive enough to detect the subtle signal currents. This contradiction between sensitivity and time
resolution is solved for pEDMR with an indirect detection scheme ~\cite{Boediss,Boe7,Boe6,Boe3,Boe9} where the change of the photocurrent after a
coherent pESR excitation is measured as a function of the length $\tau$ of the resonant pulse. A sketch of this measurement principle is
illustrated in fig.~\ref{fig:overview}: The experiment begins when the steady state ensemble $\hat\rho^S$ that, due to the short singlet
lifetimes, consists mainly of pure triplet eigenstates, is coherently manipulated and brought into a non--steady state, non--eigenstate
$\hat\rho(\tau,B_1,\omega)$ which is determined by $\tau$, the microwave field strength $B_1$ and the microwave frequency $\omega$. After the
excitation, the non--eigenstates will carry out a Larmor precession whose influence on the net transition rate will fade quickly due to the
ensemble dephasing~\cite{Boe1}. Thus, a short time after the end of the microwave pulse, a non--steady state transition rate is present that
relaxes slowly (on a $\mu$s to ms time scale) back to the steady state. It is known~\cite{Boe6} that the integral of this relaxation current, $Q$,
is proportional to the the density change
\begin{equation}
\Delta:=-\frac{\rho_{11,44}-\rho^S_{11,44}}{\mathrm{Tr}\left[\rho^S\right]}
=\frac{\rho_{22,33}-\rho^S_{22,33}}{\mathrm{Tr}\left[\rho^S\right]}\frac{\hbar\omega_\Delta}{\hbar\omega_\Delta\pm(J+D^d)},
\label{deldef1}
\end{equation}
wherein $\rho_{ii}$ and $\rho^S_{ii}$ are the density matrix and the steady state density matrix elements, respectively and $J$, $D^d$ and
$\omega_\Delta$ correspond to the exchange coupling, the dipolar coupling and the Larmor separation within the pairs,
respectively~\cite{Boe6,Boediss}. Because of this, $\Delta=\Delta\left(\hat\rho(\tau),\hat\rho^S\right)$ is a function of the ensemble state
$\hat\rho(\tau)$ right at the end of the pulse and thus, it is possible to determine the evolution of $\hat\rho(\tau)$ during the excitation by
measuring $Q$ as function of $\tau$. The time resolution of this measurement scheme is obviously not determined by the current amplifier but by
the pulse length generator and thus, a low ns--range time resolution is technically easy to achieve.

\begin{figure}[b]
\begin{center}
\includegraphics[width=84mm]{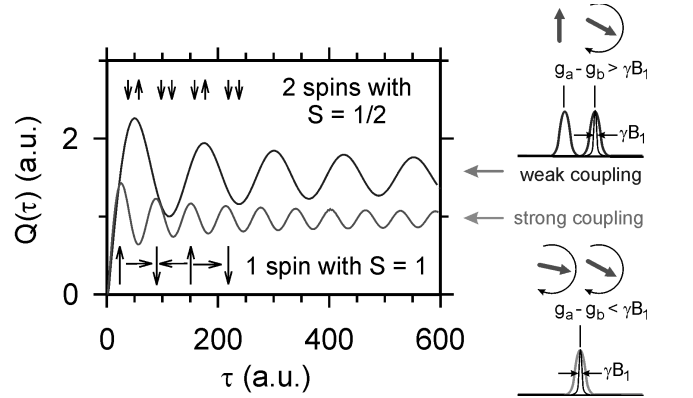}
\caption{Simulation of $Q$ as a function of $\tau$ for weak and strong spin--spin couplings. For details see text.}\label{fig:rabitheo}
\end{center}
\end{figure}

For the detection of spin--Rabi oscillation during the coherent excitation, $Q(\tau)$ can be recorded when $B_1$ is strong enough so that Rabi
frequencies are larger than the coherence time of the spin pairs and $\omega$ is in ESR with a selected defect or impurity. A quantum mechanical
description of this experiment~\cite{Boe6} has revealed an expression
\begin{equation}
\Delta\left(\tau\right)=g_i\mu_B
B_1\Phi\left(\omega\right)\int\limits_{-\infty}^{\infty}
\frac{\sin^2\left(\kappa g_i\mu_B
B_1\tau\sqrt{1+x^2}\right)}{1+x^2}dx \label{lisha}
\end{equation}
under the assumption of homogeneous $B_1$ fields and a sufficiently smooth line shape $\Phi\left(\omega\right)$ of the spins in resonance which
means $\partial_{\omega}\Phi(\omega_i)g_i\mu_B B_1\ll\Phi(\omega)$. In eq.~\ref{lisha}, pair partner $i$ has a Land\'e factor, $g_i$, and is
exposed to an external magnetic field $B_0$ whereas $\kappa$ denotes a factor whose value depends on the spin--spin coupling within a pair. An
illustration of two of these coupling cases based on a theoretical calculation~\cite{Boe6} is given in fig.~\ref{fig:rabitheo}. Weak coupling
($g_a-g_b\gg D^d,J$) implies that an ESR excitation can always manipulate either spin \textbf{a} or spin \textbf{b}, depending on the chosen
excitation frequency $\omega$. Hence, the Rabi oscillation reflects the transient nutation of a simple $s=1/2$ electron spin and therefore,
$\kappa=1/2$. When the coupling is strong ($g_a-g_b\ll D^d,J$), the excitation is not selective for any pair partner anymore and hence, two
$s=1/2$ electron spins are turned and $\kappa=1$. The transients plotted in fig.~\ref{fig:rabitheo} where calculated under negligence of
incoherence. The decay of the oscillation is due to the gradual spectral narrowing of the excitation width with increasing $\tau$.

In addition to the two coupling cases illustrated in fig.~\ref{fig:rabitheo}, another case shall be mentioned here: When $g_a-g_b\ll D^d,J$
(strong coupling) but $B_1\ll J,D^d$, then $\kappa=1/\sqrt{2}$. While this case has so far not been described theoretically for pulsed EDMR
experiments, one can deduce it from the description of transient nutation experiments of $s>1/2$ systems without hyperfine influences as given by
Astashkin and Schweiger~\cite{schweiger:1990}.

\onecolumngrid
\begin{widetext}
\begin{figure}[b]
\begin{center}
\includegraphics[width=160mm]{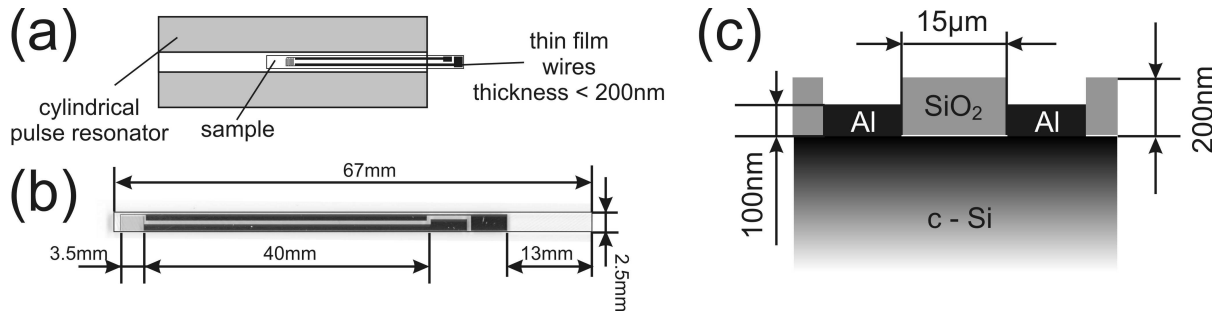}
\end{center}
\caption{(a) Sketch of a microwave mode compatible sample inserted in a dielectric microwave resonator. (b) Photo of a match like microwave mode
compatible contact structure. (c) Sketch of the sample cross section with the c-Si/$\mathrm{SiO_2}$ interface.\vspace{5mm}}\label{fig:gridsystem}
\end{figure}
\end{widetext}
\twocolumngrid

\section{A pEDMR experiment with $\mathbf{P_b}$-centers}
We have chosen recombination of photoexcited charge carriers at the well understood and well characterized $\mathrm{P_b}$-center to serve as a
model system for the demonstration of pEDMR. $\mathrm{P_b}$ centers are trivalent Si atoms at the c-Si/SiO$_2$ interface. They dominate interface
trapping and recombination, they are paramagnetic when uncharged~\cite{Poin:1981,Poin:1984} and they are strongly localized, anisotropic
electronic states~\cite{lena:1998}. For the c-Si (111) surface orientation, all $\mathrm{P_b}$ centers point into a direction perpendicular to the
interface. Because of this, their microscopic anisotropy is reflected by the ESR as well as EDMR spectra as shown repeatedly in the
literature~\cite{lena:1998,Friedrich:2004}. First pEDMR studies at the $\mathrm{P_b}$-center have been carried out recently by Friedrich et
al.~\cite{Friedrich:2004} which revealed that charge carrier trapping and recombination can take place without the presence of additional shallow
trapping centers through a two step trapping/readjustment direct capture process that had been described theoretically first by Shockley and
Read~\cite{shoc:1952} and later Rong et al.~\cite{rong2}.

In order to conduct pEDMR, a semiconductor sample must be placed inside a microwave resonator such that the $B_1$ field about the centers that are
to be excited can be generated. Since electrical contacts naturally consist of conducting material, they may alter the eigenmodes of a microwave
cavity whose geometry was designed under the assumption that the fill--factor of conducting material therein is negligible. The uncontrolled
change of eigenmodes leads to a strong inhomogeneity of $B_1$ throughout the resonator, especially at the sample position. This causes a rapid,
artificially induced dephasing of the spins in resonance and thus, the observation of Rabi oscillation becomes impossible. Thus, the sample and
especially the contacts must be designed such that a $B_1$ distortion is as small as possible. One can achieve this with sample substrates whose
conductivity is as low as possible and sample contacts with thicknesses below the microwave penetration depth. An example for such a complete thin
film contact wiring of the sample within the microwave resonator is illustrated in fig.~\ref{fig:gridsystem}(a) to (c). In (a), a sketch of a thin
film wired sample within a cylindrical microwave resonator is shown. While the actual semiconductor sample with its interdigited contact grid is
located at the tip of the match--like substrate in the center of the cavity, it is connected to the contact pads on the outside by 40 mm long and
less than 200 nm thin Al stripes. Figure~\ref{fig:gridsystem}(b) displays a photo of an acutal thin--film wire and contact structure. One can see
the structure and its dimensions with contact pads, wires and grids. The grid area consists of 75 grid pairs where each grid has 5 $\mu$m width
and 15 $\mu$m distance to its respective neighbors as indicated in fig.~\ref{fig:gridsystem}(c). With the sample and contact geometry given, one
can (i) minimize sample resistances and therefore maximize sample currents, time resolution and sensitivity and (ii) the actual semiconductor
sample will be at the center of the cavity where $B_1$ has its maximum, while (iii) the eigenmodes of the cavity especially at its center remain
undistorted.

The sample used for the experiment was made from a 380 $\mu$m thick (111) surface oriented, slightly  phosphorous doped ($\approx$ 10 $\Omega$cm)
Czochralski grown c-Si wafer whose surface was subjected to an RCA cleaning procedure followed by the formation of an about 200 nm thick thermal
oxide layer. The oxide was formed at 1050 $^o$C under exposure of the sample to O$_2$ for 200 min. The thickness of the resulting oxide was
confirmed by profilometer measurements. After the oxide formation, the contact and wire system was deposited by means of a photolithographic
lift--off procedure before the wafer was cut into match--like stripes. The current detection and coherent microwave excitation as well as the
extraction of spin--dependent currents from microwave induced currents was executed with the same setup and the same procedure as described in
Refs.~\cite{Boediss} and \cite{Boe7}.

All experiments presented in the following were performed at a sample temperature of $T=10$ K and a sample
\begin{figure}[t]
\begin{center}
\includegraphics[width=84mm]{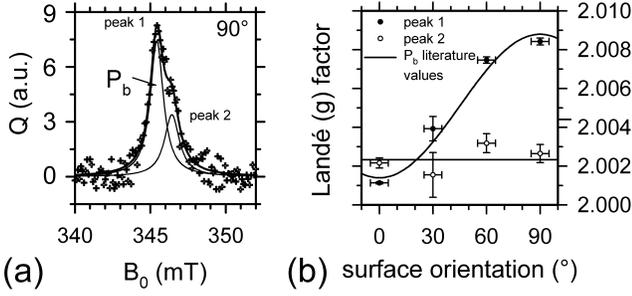}
\end{center}
\caption{(a) Measurement of $Q(B_0)$ for a sample orientation of 90$^\circ$ and a fit with two Lorentzian lines (solid line). (b) The $g$ factors
of the two peaks observed in (a) as a function of the sample orientation with respect to $B_0$. } \label{fig:Pb_ident}
\end{figure}
irradiation of 0.2(1) W/cm$^2$ with Ar$^+$ laser light ($\lambda=514$ nm). For the coherent excitation, a Bruker E580 X-Band pulse ESR
spectrometer was used. A photocurrent of $I=5$ $\mu$A was established by a constant current source with long dwell time (1 s) to allow for a drift
compensation. In order to measure the charge $Q$, the current was then subtracted by the constant offset before it was transformed by an impedance
changer into a voltage signal which was then filtered by a high pass and subsequently digitized by an 8 bit transient recorder. The integration
took place between 14 $\mu$s and 30 $\mu$s after the pulse which is the part of the photocurrent relaxation transient where the current signal
reached its maximum.

The signal to noise ratios (SNR) per charge carrier pair, which poses an upper limit for the spin sensitivity since several charge carrier pairs
can undergo transitions at one defect, was $\frac{SNR}{\mathrm{eh-pair}}=\frac{\sqrt{n}}{10^6}$ at 250 W microwave power with n being the number
of accumulated transients. Thus, within an 8 hour period and 300 $\mu$s shot repetition time, one can attain a sensitivity of a few hundred charge
carriers. This sensitivity is about 9 orders of magnitude higher than conventional ESR measurements under comparable conditions, yet it is still
not a single spin or even a single spin per single shot sensitivity. The reason for this limit is the presence of strong artifact currents within
the sample due to the application of the strong microwave pulses. It is therefore not a principle limitation of the measurement method but due to
the sample design. A further downscaling of the sample area and the prevention of shunt currents due to diffused excess charge carriers into the
c-Si bulk could push the sensitivity even further.
\begin{figure}[b]
\begin{center}
\includegraphics[width=84mm]{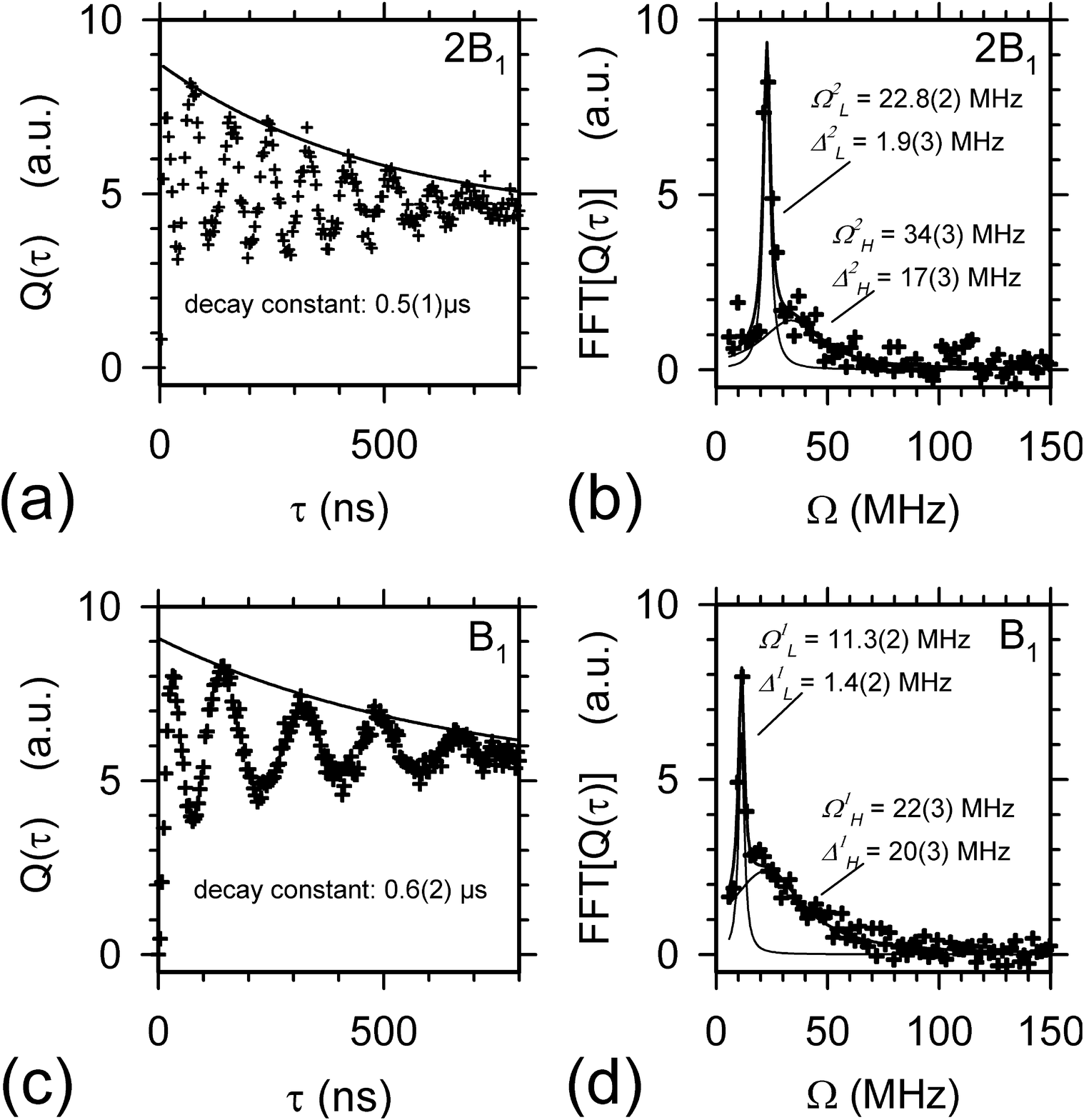}
\end{center}
\caption{(a) and (c): Measurement of $Q(\tau)$ for a sample orientation of 90$^\circ$ for two arbitrary microwave fields $2B_1$ and $B_1$,
respectively. (b) and (d): The fast Fourier transform of the data in (a) and (c), respectively. The nomenclature of the fit results is defined in
the figure and referred to in the text.} \label{fig:Rabi_decay}
\end{figure}

\section{Experimental results}
We have recorded $Q$ as a function of the strength of the external magnetic field $B_0$ as well as the angle between $B_0$ and the c-Si (111)
orientation of the sample in order to confirm that the measured signals are due to electronic transitions at $\mathrm{P_b}$ centers.
Figure~\ref{fig:Pb_ident}(a) displays this for a sample orientation of 90$^\circ$ which was recorded after an excitation with $\tau=400$ ns and a
microwave power of 4 W. The data was fit with two Lorentzian line shapes. One can see that peak 1 has a much stronger intensity in comparison to
peak 2. The measurement represented by fig.~\ref{fig:Pb_ident}(a) was repeated for sample orientations with angles of 60$^\circ$, 30$^\circ$, and
0$^\circ$. In all cases, the data could be fit reasonably with two Lorentzians. The $g$ factor of all fits are displayed in
fig.~\ref{fig:Pb_ident}(b) and show that peak 1 has the anisotropy of the $\mathrm{P_b}$ centers as it can be found in the
literature~\cite{lena:1998} and as it is shown by the solid line. We therefore assign peak 1 to the $\mathrm{P_b}$ center. Peak 2 shows no
identifyable anisotropy.

For the electrically detected spin--Rabi oscillations, the sample orientation was turned back into the 90$^\circ$ position since then, the two
peaks were well separated such that an excitation of peak 2 was minimized. The microwave frequency and the external field $B_0$ where adjusted
such that the peak maximum at $g\approx2.008$ was on resonance. Then, $\tau$ was changed between 0 and 800 ns in 2 ns steps. The resulting
transients are displayed in fig.~\ref{fig:Rabi_decay}(a) and (c) for applied microwave powers of 250W and 62W, respectively. This corresponds to
arbitrary microwave fields $2B_1$ and $B_1$, respectively. One can clearly see the oscillatory behavior of $Q$ in both plots.

The maxima in the plots (a) and (c) of fig.~\ref{fig:Rabi_decay} were fit with simple exponential decay functions. The two fits agree within the
margin of error and reveal a time constant of $\approx 500$ ns. In order to determine the frequency components of the two data sets, they were
subjected to fast Fourier transforms (FFT) whose absolute results are plotted in fig.~\ref{fig:Rabi_decay}(b) and (d) for the microwave fields
$2B_1$ and $B_1$, respectively. The data of both plots was fit with two Lorentzians with the fit plotted in the graphs. Note that, within the
margin of error, the exponential decay functions of fig.~\ref{fig:Rabi_decay}(a) and (c) agree with the fit results for the width of the peaks
with lower frequency $\Delta^1_L$ and $\Delta^2_L$ in the FFT plots (b) and (d) which shows that the lower frequency component belongs to the
slowly decaying process.

\section{Discussion of results}
We have shown that it is possible to observe coherent spin--Rabi oscillation by means of transient current measurements. This shows that distinct
eigenstates of spins in semiconductors can be read electrically. One can interpret the data presented ins figs.~\ref{fig:Pb_ident} and
\ref{fig:Rabi_decay} by taking advantage of Rabis formula $\Omega=\sqrt{\kappa B_1^2+(\omega-\omega_i)^2}$~\cite{Boe6}, wherein $B_1$ is expressed
in gyromagnetic units. Note that $\Omega\propto B_1$ when we measure on resonance. Since this is exactly the case for the center frequencies
$\Omega^1_L$ and $\Omega^2_L$ of peak 1 as shown in fig.~\ref{fig:Rabi_decay}(b) and (d), one can conclude that the slowly decaying oscillation is
solely due to spin--Rabi oscillation involving the $\mathrm{P_b}$ center. For the higher, broadly distributed frequencies centered around
$\Omega^1_H$ and $\Omega^2_H$ this is different: From Rabis formula, we learn that when we measure off--resonant at arbitrary Rabi frequencies
$\Omega^1$ and $\Omega^\xi$ with two different $B_1$ fields with ratio $B_1^1=\xi B_1^\xi$, the Larmor--frequency difference can be calculated as
\begin{equation}
\omega_i-\omega=\sqrt{\frac{\xi^2{\Omega^1}^2-{\Omega^\xi}^2}{\xi^2-1}}.
\label{eqf}
\end{equation}
With regard to the data in fig.~\ref{fig:Rabi_decay} where $\xi=2$, this means that $\omega-\omega_i=16(10)\mathrm{MHz}\simeq0.6(4)$mT for the
broad peaks in fig.~\ref{fig:Rabi_decay}(b) and (d) and thus, we can conclude, that the process responsible for these peaks can be associated with
peak 2 of fig.~\ref{fig:Pb_ident}(a) whose Larmor--frequency $\omega_2$ was about 0.9mT higher than the excitation frequency $\omega$.

Another consequence of Rabis formula mentioned above is that when we have different frequency components $\Omega_H$ and $\Omega_L$ in one
measurement, we can obtain the ratio of their coupling factors
\begin{equation}
\frac{\kappa_H}{\kappa_L}=\sqrt{\frac{{\Omega^1_H}^2-{\Omega^2_H}^2}{{\Omega^1_L}^2-{\Omega^2_L}^2}}
\label{eqf2}
\end{equation}
from two measurements 1 and 2 collected at two arbitrary but different $B_1$ fields. Note that eq.~\ref{eqf2} is independent of the
Larmor--frequency difference which means it does not play a role whether the pair centers are on or off resonance. Thus, when the fit results of
fig.~\ref{fig:Rabi_decay}(b) and (d) are plugged into eq.~\ref{eqf2} we obtain $\frac{\kappa_H}{\kappa_L}=1.3(3)$. With the margin of error for
$\frac{\kappa_H}{\kappa_L}$ given, one can not distinguish whether $\frac{\kappa_H}{\kappa_L}=1$ or $\frac{\kappa_H}{\kappa_L}=\sqrt{2}$. However,
since the $\mathrm{P_b}$ mechanism associated with peak 1 has been attributed to a direct capture process into strongly coupled
$\mathrm{P_b^{-*}}$ pairs in the past~\cite{Friedrich:2004}, it becomes clear that the process associated with peak 2 must involve strongly
coupled pairs, too.

\section{Summary}
The ultra--sensitive electrical detection of electron spin--Rabi oscillation at $\mathrm{P_b}$ centers has been demonstrated and it was shown how
new insights into the nature of charge carrier recombination at $\mathrm{P_b}$ centers can be gained. In addition, PEDMR at the c-Si/SiO$_2$
interface showed that a second, from the $\mathrm{P_b}$--direct capture process qualitatively different spin--dependent transition exits. With the
used sample design, the electrical measurements of coherent spin motion reached a sensitivity below 1000 spins. Since this limitation is purely
due to offset and microwave artifact currents, it is conceivable that further downscaling and a different sample design may be able to shift this
limit towards a coherent single spin readout.

\section{Acknowledgements}
We thank Kerstin Jacob for the sample preparation and Jan Behrends, Kai Petter, Felice Friedrich, Walther Fuhs and Gerry Lucovsky for inspiring
discussions


\begin{thebibliography}{}
\expandafter\ifx\csname url\endcsname\relax
  \def\url#1{\texttt{#1}}\fi
\expandafter\ifx\csname
urlprefix\endcsname\relax\def\urlprefix{URL }\fi

\bibitem{Rugar:2004}
{D. Rugar} and {R. Budakian} and {H. J. Mamin} and {B. W. Chui}, Nature (London) 430 (2004) 329.
\bibitem{Jelezko:2002}
{F. Jelezko} and {I. Popa} and {A. Gruber} and {C. Tiez} and {J. Wrachtrup} and {A. Nizovtsev} and {S. Kilin}, Appl. Phys. Lett. 81~(12) (2002)
2160.
\bibitem{Jelezko:2004}
{F. Jelezko} and {T. Gaebel} and {I. Popa} and {A. Gruber} and {J. Wrachtrup}, Phys. Rev. Lett. 92~(7) (2004) 076401--1.
\bibitem{Kouven:2004}
{J. M. Elzerman} and {R. Hanson} and {L. H. Willems van Beveren} and {B. Witkamp} and {L. M. K. Vandersypen} and {L. P. Kouwenhoven}, Nature
(London) 430 (2004) 431.
\bibitem{Jiang:2004}
{M. Xia} and {I. Martin} and {E. Yablonovitch} and {H. W. Jiang}, Nature (London) 430 (2004) 435.
\bibitem{Boediss}
{Christoph B\"ohme}, Dynamics of spin--dependent charge carrier recombination, Cuvillier Verlag, {G\"ottingen}, 2003.
\bibitem{Boe7}
{C. Boehme} and {K. Lips}, Phys. Rev. Lett. 91~(24) (2003) 246603.
\bibitem{KSM}
{D. Kaplan} and {I. Solomon} and {N. F. Mott}, J. Phys. (Paris) -- Lettres 39~(4) (1978) L51--L54.
\bibitem{HabDie}
{R. Haberkorn} and {W. Dietz}, Solid State Commun. 35~(6) (1980) 505--508.
\bibitem{Boe6}
{C. Boehme} and {K. Lips}, Phys. Rev. B 68~(24) (2003) 245105.
\bibitem{Boe3}
{C. Boehme} and {K. Lips}, Appl. Phys. Lett. 79~(26) (2001)
4363--4365.
\bibitem{Boe9}
{C. Boehme} and {K. Lips}, Appl. Magnetic Resonance 27 (2004) 109--122.
\bibitem{Boe1}
{C. Boehme} and {P. Kanschat} and {K. Lips}, Europhys. Lett. 56~(5) (2001) 716--721.
\bibitem{schweiger:1990}
{A. V. Astashkin} and {A. Schweiger}, Chemical Physics Letters 174~(6) (1990) 595--602.
\bibitem{Poin:1981}
E. H. Poindexter and P. J. Caplan and B. E. Deal and R. R. Razouk, J. Appl. Phys. 52 (1981) 879.
\bibitem{Poin:1984}
E. H. Poindexter and G. J. Gerardi and M.-E. Rueckel and P. J. Caplan and N. M. Johnson and D. K. Biegelsen, J. Appl. Phys. 56~(10) (1984) 2844.
\bibitem{lena:1998}
P. M. Lenahan and J. F. Conley, J. Vac. Sci. Technol. B 16 (1998) 2134--2153.
\bibitem{Friedrich:2004}
{F. Friedrich} and {C. Boehme} and {K. Lips}, J. Appl. Phys. 97 (2005) 056101.
\bibitem{shoc:1952}
W. Shockley and W. T. Read, Jr., Phys. Rev. 87~(5) (1952) 835.
\bibitem{rong2}
{F. C. Rong} and {G. J. Gerardi} and {W. R. Buchwald} and {E.H. Poindexter} and {M. T. Umlor} and {D. J. Keeble} and {W. L. Warren}, Appl. Phys.
Lett. 60~(5) (1992) 610--612.
\end{thebibliography}
\end{document}